\documentclass[conference]{IEEEtran}
\IEEEoverridecommandlockouts

\usepackage{cite}
\usepackage{amsmath,amssymb,amsfonts}
\usepackage{algorithmic}
\usepackage{graphicx}
\usepackage{textcomp}
\usepackage{xcolor}
\usepackage{booktabs} 
\usepackage{qcircuit}

\def\BibTeX{{\rm B\kern-.05em{\sc i\kern-.025em b}\kern-.08em
    T\kern-.1667em\lower.7ex\hbox{E}\kern-.125emX}}
\begin{document}

\title{Quantum Reinforcement Learning Trading Agent for Sector Rotation in the Taiwan Stock Market
}

\author{
\IEEEauthorblockN{Chi-Sheng Chen}
\IEEEauthorblockA{\textit{Neuro Industry Research} \\
\textit{Neuro Industry, Inc.}\\
Cambridge, Massachusetts, USA \\
m50816m50816@gmail.com}
\and
\IEEEauthorblockN{Xinyu Zhang}
\IEEEauthorblockA{\textit{Department of Computer Science} \\
\textit{The University of Alabama}\\
Tuscaloosa, Alabama, USA \\
xzhang205@crimson.ua.edu}
\and
\IEEEauthorblockN{Ya-Chuan Chen}
\IEEEauthorblockA{\textit{Department of Information and}\\
\textit{Communications Research Laboratories} \\
\textit{Industrial Technology Research Institute}\\
Hsinchu, Taiwan \\
joycechen108@gmail.com}
}

\maketitle

\begin{abstract}
We propose a hybrid quantum-classical reinforcement learning framework for sector rotation in the Taiwan stock market. Our system employs Proximal Policy Optimization (PPO) as the backbone algorithm and integrates both classical architectures (LSTM, Transformer) and quantum-enhanced models (QNN, QRWKV, QASA) as policy and value networks. An automated feature engineering pipeline extracts financial indicators from capital share data to ensure consistent model input across all configurations. Empirical backtesting reveals a key finding: although quantum-enhanced models consistently achieve higher training rewards, they underperform classical models in real-world investment metrics such as cumulative return and Sharpe ratio. This discrepancy highlights a core challenge in applying reinforcement learning to financial domains—namely, the mismatch between proxy reward signals and true investment objectives. Our analysis suggests that current reward designs may incentivize overfitting to short-term volatility rather than optimizing risk-adjusted returns. This issue is compounded by the inherent expressiveness and optimization instability of quantum circuits under Noisy Intermediate-Scale Quantum (NISQ) constraints. We discuss the implications of this reward-performance gap and propose directions for future improvement, including reward shaping, model regularization, and validation-based early stopping. Our work offers a reproducible benchmark and critical insights into the practical challenges of deploying quantum reinforcement learning in real-world finance.
\end{abstract}


\section{Introduction}
Financial markets are inherently dynamic and noisy~\cite{chiarella2011analysis}, driven by complex interactions among macroeconomic cycles, sector-specific trends, and investor behaviors. Sector rotation, the practice of reallocating capital between industry sectors based on their expected performance across economic phases, is a prominent strategy for generating alpha and managing risk in volatile markets. Reinforcement learning (RL)~\cite{kaelbling1996reinforcement} offers a powerful paradigm for this challenge. Its ability to learn optimal decision policies through direct interaction with an environment aligns naturally with the sequential, stochastic nature of financial decision-making. Unlike traditional rule-based or supervised learning methods that depend on static datasets, RL agents dynamically adapt their strategies to maximize cumulative rewards over time.

Recent advances in quantum machine learning (QML)~\cite{biamonte2017quantum} introduce new possibilities for enhancing these learning systems. By integrating quantum principles like superposition and entanglement, quantum reinforcement learning (QRL)~\cite{dong2008quantum, chen2022variational} aims to create more expressive policy representations and accelerate optimization. However, the practical application of QRL is currently constrained by the limitations of Noisy Intermediate-Scale Quantum (NISQ) hardware, including significant noise, shallow circuit depth, and barren plateau phenomena~\cite{chen2020variational, kuo2021quantum}. In this work, we investigate the application of a hybrid quantum-classical RL framework to the task of sector rotation within the Taiwan stock market. We have developed a flexible agent based on Proximal Policy Optimization (PPO)~\cite{schulman2017proximal} that supports both classical network backbones, such as LSTM~\cite{hochreiter1997long} and Transformer~\cite{vaswani2017attention}, and quantum-enhanced backbones, including QNN~\cite{mitarai2018quantum}, QRWKV~\cite{chen2025qrwkv}, and QASA~\cite{chen2025quantum}. All agents were trained and backtested under consistent experimental conditions using sector-level capital share data and automated feature engineering. A key finding of our study is that while quantum models frequently achieved higher rewards during training, they consistently underperformed classical models on real-world investment metrics like cumulative return and the Sharpe ratio. This highlights a fundamental disconnect between proxy reward functions and actual investment goals in financial RL applications. Furthermore, it raises critical questions about the interplay between quantum expressiveness, training instability, and reward design within the constraints of NISQ-era technology.

Our contributions are threefold. First, we introduce a reproducible QRL benchmark for sector rotation using real financial data. Second, we provide a systematic comparison of classical and quantum-enhanced policy networks within a shared PPO framework. Third, we analyze the observed discrepancy between training rewards and investment performance, proposing actionable improvements for reward shaping and regularization to bridge this gap.

\begin{figure}
    \centering
    \includegraphics[width=1\linewidth]{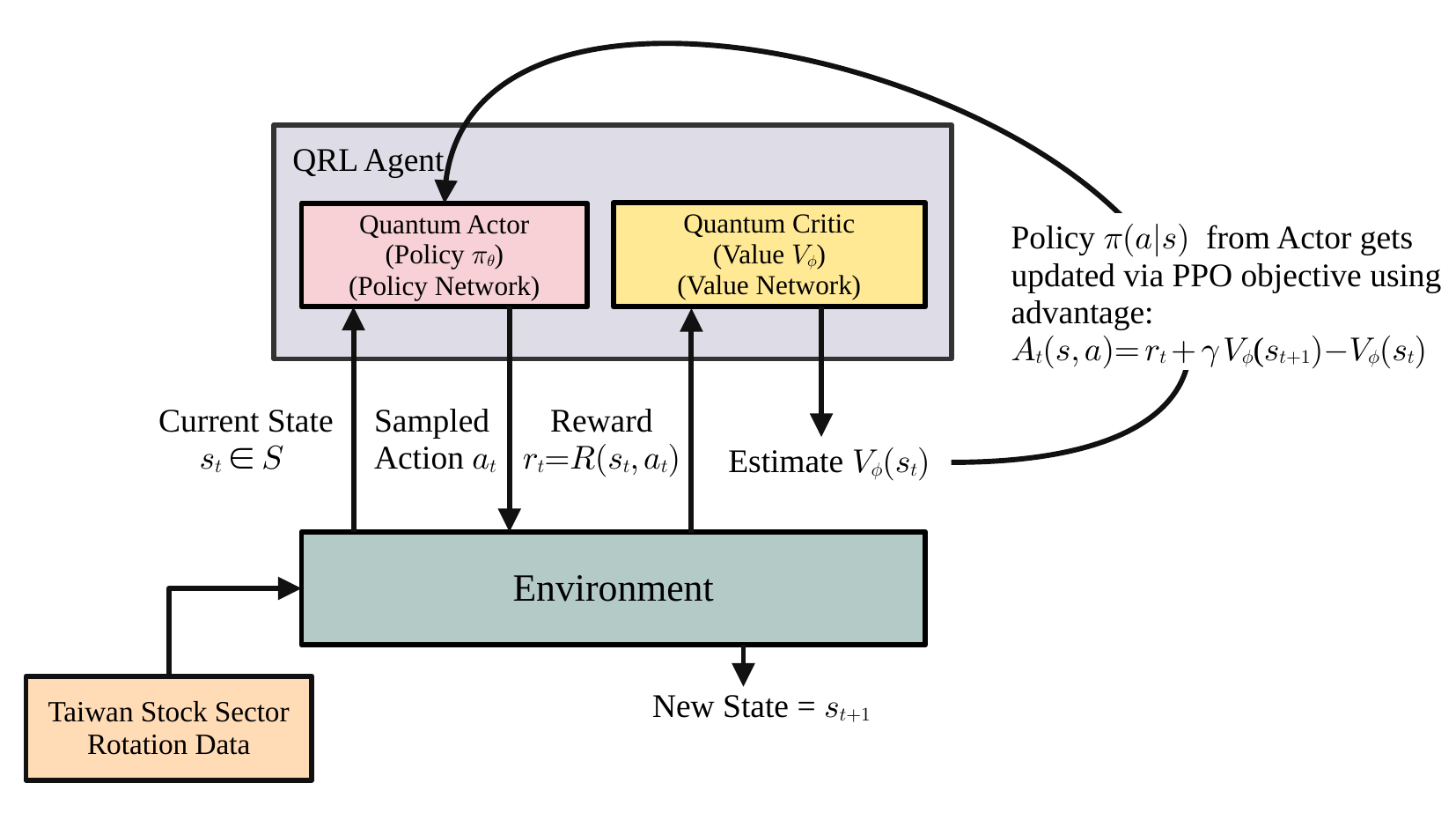}
    \caption{The overview of this work.}
    \label{fig:qrl_ppo}
\end{figure}

\section{Related Work}

 \subsection{Sector Rotation Strategies in Financial Markets}

Sector rotation is a dynamic and adaptive investment strategy that systematically reallocates capital among various industry sectors based on expectations of future economic performance~\cite{cavaglia2001risks}. The underlying principle assumes that different sectors respond differently to various phases of the economic cycle. For example, technology and consumer discretionary sectors often outperform during economic expansions, while defensive sectors such as utilities and consumer staples tend to be more resilient during contractions. Conventional approaches to sector rotation can be categorized into top-down and bottom-up methodologies. The top-down approach involves analyzing macroeconomic indicators such as gross domestic product growth, interest rates, and inflation to infer sector performance trends. In contrast, the bottom-up approach focuses on evaluating the fundamentals of individual companies within each sector to identify opportunities. A combined approach seeks to synthesize both macroeconomic and microeconomic perspectives to form a holistic investment strategy. More recently, quantitative techniques have been employed to enhance sector rotation, including factor-based models that exploit patterns such as momentum, mean reversion, and volatility clustering. However, many of these methods remain static or rule-based, lacking the ability to adapt autonomously to new market conditions or structural changes. This motivates the exploration of reinforcement learning as a means of enabling adaptive, data-driven sector allocation strategies.

\subsection{Reinforcement Learning in Portfolio Management}

Reinforcement learning (RL) provides a computational framework in which an agent learns to make sequential decisions through interactions with a dynamic environment. In the context of financial markets, RL has been applied to various tasks such as portfolio rebalancing, asset allocation, algorithmic trading, and hedging strategies. By learning policies that maximize long-term cumulative rewards, RL agents are capable of uncovering investment strategies that adapt over time. Compared to traditional approaches such as the Markowitz mean-variance optimization framework, RL does not rely on strong statistical assumptions about asset return distributions or covariance matrices. This makes RL particularly well-suited for financial environments, which are often characterized by high volatility, structural breaks, and non-stationary behavior. Furthermore, the incorporation of deep neural networks in deep reinforcement learning (DRL) enables the modeling of nonlinear and temporally dependent financial patterns. Notably, architectures such as LSTM networks and Transformer encoders have been used to capture complex dependencies in sequential financial data, improving an agent’s ability to anticipate market dynamics. However, despite these advances, one of the persistent challenges in applying RL to financial decision-making lies in the design of appropriate reward functions. Agents may overfit to proxy rewards that correlate poorly with true investment objectives, such as risk-adjusted returns, resulting in strategies that perform well during training but fail during backtesting or live trading.

\subsection{Quantum Machine Learning for Financial Optimization}

Quantum machine learning (QML) is an emerging field that explores the integration of quantum computation with machine learning algorithms. By harnessing quantum mechanical phenomena such as superposition, entanglement, and quantum parallelism, QML aims to improve learning efficiency, representational power, and solution quality, especially for high-dimensional or computationally intractable problems. In finance, QML has been proposed for a wide range of applications, including portfolio optimization, option pricing, credit risk modeling, and market simulation. Hybrid quantum-classical models, which combine variational quantum circuits with classical neural networks, have become a practical choice for current hardware capabilities, particularly in the Noisy Intermediate-Scale Quantum (NISQ) era. These models seek to leverage quantum expressiveness while maintaining the stability and scalability of classical computation. Quantum reinforcement learning (QRL), a subfield of QML, focuses on embedding quantum components within RL architectures, such as representing policies or value functions using quantum circuits. Several studies have demonstrated the theoretical benefits of QRL in synthetic environments, showing potential improvements in learning capacity and sample efficiency. Nevertheless, empirical evaluations on real-world tasks remain limited. Key challenges include the difficulty of training quantum models due to barren plateau phenomena, susceptibility to noise and decoherence, and the absence of standardized quantum-compatible financial environments. This work contributes to the growing body of literature by presenting a practical and reproducible QRL framework applied to the real-world problem of sector rotation. In contrast to prior research that primarily focuses on proof-of-concept simulations, our study offers a direct comparison between classical and quantum-enhanced models under consistent financial conditions and highlights the mismatch between training reward optimization and actual investment performance.

\section{Methodology}

\subsection{Problem Formulation}

We formulate the sector rotation problem as a sequential decision-making task in which an agent selects an allocation across multiple industry sectors to maximize long-term investment return. At each discrete time step $t$, the agent observes a financial state vector $s_t$, derived from engineered features of sector-level market data. Based on this state, the agent chooses an action $a_t$, representing the target portfolio allocation across sectors. After executing the action, the environment transitions to a new state $s_{t+1}$ and returns a scalar reward $r_t$ based on the investment outcome. The goal is to learn a policy $\pi(a_t|s_t)$ that maximizes the expected cumulative discounted reward $\mathbb{E}\left[\sum_t \gamma^t r_t\right]$.

\subsection{Framework Overview}

We adopt a hybrid quantum-classical reinforcement learning framework based on the Proximal Policy Optimization (PPO) algorithm. PPO is a widely used policy-gradient method known for its sample efficiency and training stability. It optimizes a clipped surrogate objective to prevent large policy updates and incorporates entropy regularization to encourage exploration. Our framework supports interchangeable policy and value network backbones, allowing both classical and quantum-enhanced models to be evaluated under consistent training protocols.

\subsection{Policy Network Architectures}

\subsubsection{Classical Baselines}

We implement two classical neural network architectures as baselines:

\begin{itemize}
    \item \textbf{LSTM~\cite{hochreiter1997long}:} The Long Short-Term Memory (LSTM) network is a type of recurrent neural network (RNN) designed to capture long-term dependencies in sequential data. For an input sequence $X = \{x_1, x_2, \ldots, x_T\}$, the LSTM updates its hidden state $h_t$ and cell state $c_t$ at each time step $t$ using the following equations:

    \begin{align*}
        f_t &= \sigma(W_f x_t + U_f h_{t-1} + b_f) \quad \text{(forget gate)} \\
        i_t &= \sigma(W_i x_t + U_i h_{t-1} + b_i) \quad \text{(input gate)} \\
        o_t &= \sigma(W_o x_t + U_o h_{t-1} + b_o) \quad \text{(output gate)} \\
        \tilde{c}_t &= \tanh(W_c x_t + U_c h_{t-1} + b_c) \quad \text{(cell candidate)} \\
        c_t &= f_t \odot c_{t-1} + i_t \odot \tilde{c}_t \quad \text{(cell state update)} \\
        h_t &= o_t \odot \tanh(c_t) \quad \text{(hidden state)}
    \end{align*}

    where $\sigma$ is the sigmoid function, $\odot$ denotes element-wise multiplication, and $W_*$, $U_*$, $b_*$ are learnable parameters.

    \item \textbf{Transformer~\cite{vaswani2017attention}:} The Transformer model relies entirely on attention mechanisms to model sequential data, without recurrence. Given an input sequence $X \in \mathbb{R}^{T \times d}$, each position attends to all others via scaled dot-product attention. The attention for a query $Q$, key $K$, and value $V$ is computed as:

    \begin{align*}
        \text{Attention}(Q, K, V) = \text{softmax} \left( \frac{QK^\top}{\sqrt{d_k}} \right)V
    \end{align*}

    where $Q = XW^Q$, $K = XW^K$, $V = XW^V$ are learned projections and $d_k$ is the dimension of the key vectors. Multi-head attention extends this by running $h$ attention heads in parallel:

    \begin{align*}
        \text{MultiHead}(X) &= \text{Concat}(\text{head}_1, \dots, \text{head}_h)W^O \\
        \text{head}_i &= \text{Attention}(Q_i, K_i, V_i)
    \end{align*}

    After attention, each layer applies a position-wise feed-forward network (FFN) and residual connections:

    \begin{align*}
        \text{FFN}(x) = \max(0, xW_1 + b_1)W_2 + b_2
    \end{align*}

    The final output is passed through layer normalization and stacked over multiple layers to form deep representations.
\end{itemize}

\subsubsection{Quantum-Enhanced Variants}

We implement three quantum-enhanced variants by incorporating variational quantum circuits (VQCs) into the policy and/or value networks:
\begin{figure}[h]
\centering
\[
\Qcircuit @C=1em @R=.7em {
\lstick{|0\rangle} & \gate{R_x(x_0)} & \gate{R_z(x_0)} & \gate{R_y(\theta_{0})} & \ctrl{1} & \qw \\
\lstick{|0\rangle} & \gate{R_x(x_1)} & \gate{R_z(x_1)} & \gate{R_y(\theta_{1})} & \targ & \qw \\
\lstick{\vdots} & \vdots & \vdots & \vdots & & \\
\lstick{|0\rangle} & \gate{R_x(x_{n-1})} & \gate{R_z(x_{n-1})} & \gate{R_y(\theta_{n-1})} & \qw & \qw \\
}
\]
\caption{QNN quantum circuit with angle embedding and entanglement}
\end{figure}
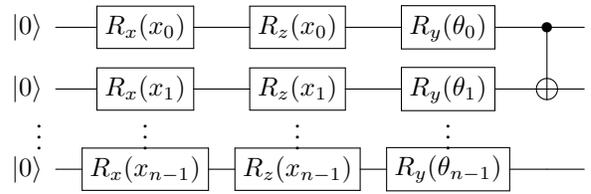

\begin{figure}
    \centering
    \includegraphics[width=1\linewidth]{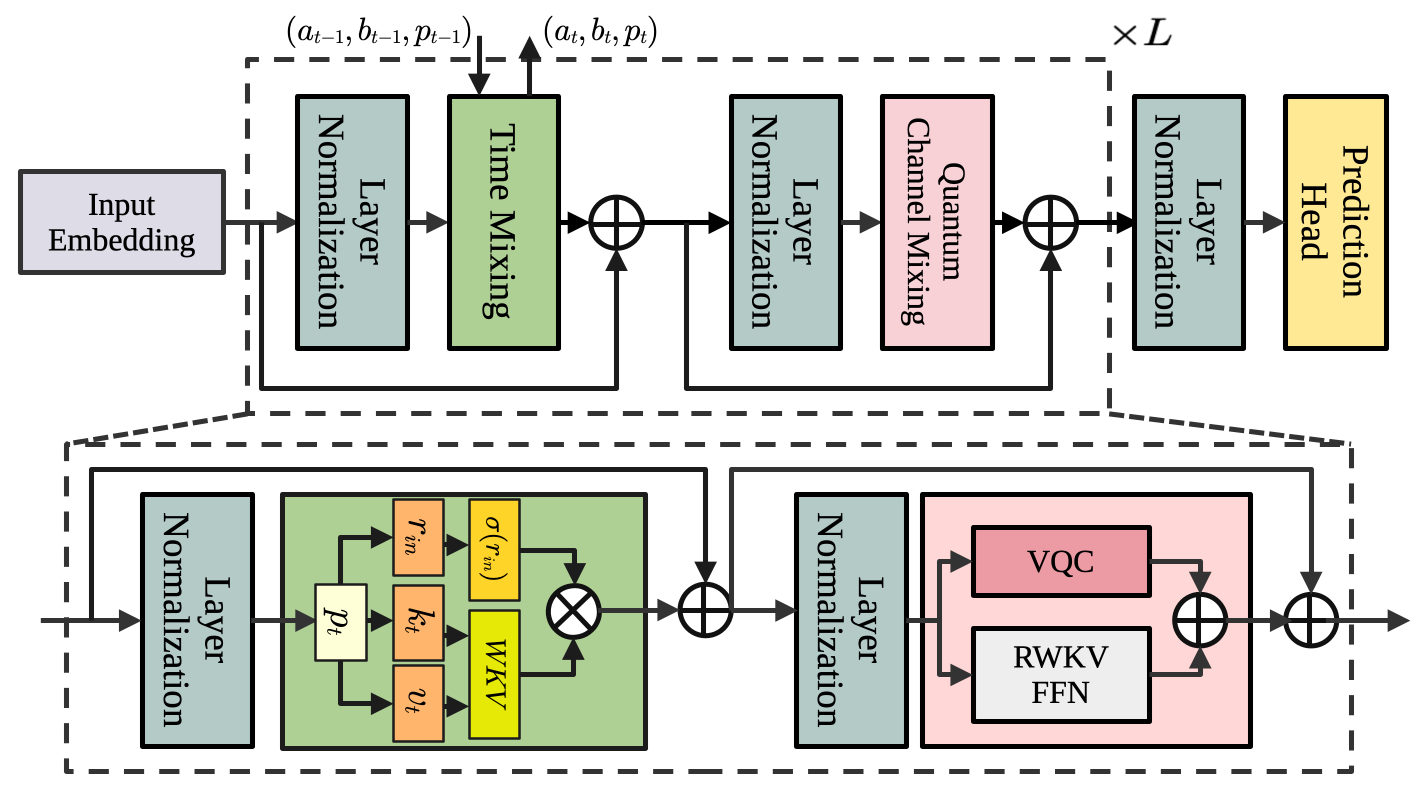}
    \caption{Architecture of a single QuantumRWKV layer.}
    \label{fig:qrwkv}
\end{figure}

\begin{itemize}
    \item \textbf{QNN~\cite{mitarai2018quantum}:} A traditional quantum neural network that projects classical input $x \in \mathbb{R}^{d}$ into a low-dimensional quantum state using angle embedding. Given a trainable projection $x' = W x \in \mathbb{R}^{n}$ with $n$ equal to the number of qubits, we apply an embedding via single-qubit rotations:
    \begin{align*}
        \text{Embed:} \quad &|0\rangle^{\otimes n} \xrightarrow{\text{RX}(x'_i), \text{RZ}(x'_i)} |\psi_{\text{in}}(x)\rangle \\
        \text{VQC:} \quad &|\psi_{\text{in}}(x)\rangle \xrightarrow{\text{U}(\theta)} |\psi_{\text{out}}(x, \theta)\rangle
    \end{align*}
    The circuit $\text{U}(\theta)$ consists of entangling layers (e.g., CNOT gates) and parameterized rotations. The measurement is performed as:
    \[
        z_i = \langle \psi_{\text{out}} | Z_i | \psi_{\text{out}} \rangle, \quad \text{for } i = 1, \dots, n
    \]
    The output vector $z$ is treated as a latent feature for downstream prediction.

    \item \textbf{QRWKV~\cite{chen2025qrwkv}:} A quantum RWKV model that replaces part of the feedforward channel mixing layer with a variational quantum circuit. For input $x \in \mathbb{R}^{d}$, the classical projection $x' = W x$ is mapped to quantum observables using:
    \[
        z = \text{QuantumCircuit}(x') \in \mathbb{R}^{n}
    \]
    The RWKV recurrence is preserved, and quantum layers enhance expressiveness in the time-mixing component:
    \[
        h_t = \text{Mix}(h_{t-1}, x_t; \theta_{\text{RWKV}}) + \text{QVC}(x_t; \theta_q)
    \]

    \item \textbf{QASA \cite{chen2025quantum}:} A quantum attention variant of the Transformer, where the classical dot-product attention:
    \[
        \text{Attention}(Q, K, V) = \text{softmax}\left( \frac{Q K^\top}{\sqrt{d}} \right)V
    \]
    is replaced by quantum attention computed via a variational quantum circuit. Let $q_i, k_j \in \mathbb{R}^{d}$ be token queries and keys, the attention score is computed as:
    \[
        \alpha_{ij} = \text{QuantumAttention}(q_i, k_j; \theta_q) \in [0, 1]
    \]
    where the circuit encodes $(q_i, k_j)$ into quantum states and returns expectation values approximating a similarity measure. The value aggregation is:
    \[
        \text{QASA}(q_i, K, V) = \sum_j \alpha_{ij} v_j
    \]
    making the attention learnable in quantum Hilbert space, enhancing expressiveness with non-linear correlations between tokens.
\end{itemize}


\begin{figure}
    \centering
    \includegraphics[width=1\linewidth]{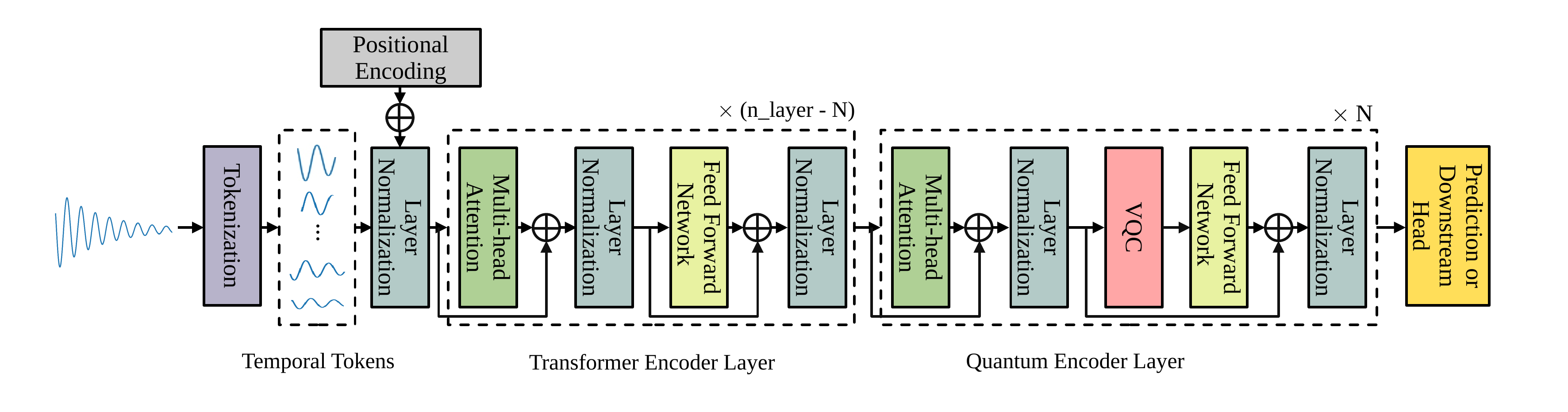}
    \caption{QASA model Architecture.}
    \label{fig:qasa}
\end{figure}

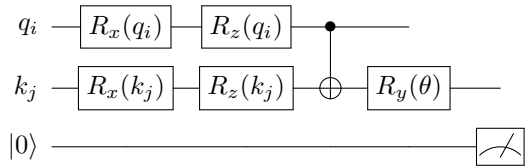
\begin{figure}[h]
\centering
\[
\Qcircuit @C=1em @R=.7em {
\lstick{q_i} & \gate{R_x(q_i)} & \gate{R_z(q_i)} & \ctrl{1} & \qw \\
\lstick{k_j} & \gate{R_x(k_j)} & \gate{R_z(k_j)} & \targ & \gate{R_y(\theta)} & \qw \\
\lstick{|0\rangle} & \qw & \qw & \qw & \qw & \meter
}
\]
\caption{QASA quantum attention: query-key interaction via entanglement and measurement}
\end{figure}
 
All quantum models use angle embedding to encode classical features into quantum states, and the output of quantum circuits is post-processed via measurement and integrated into classical layers. The design of these quantum components reflects trade-offs between expressivity and trainability under NISQ constraints.

\subsection{Feature Engineering}

To ensure a consistent and expressive input representation across all models, we construct a unified feature matrix $\mathbf{X} \in \mathbb{R}^{T \times d}$ from sector-level market data using automated feature engineering. Each feature vector $\mathbf{x}_t \in \mathbb{R}^{d}$ at time step $t$ consists of a concatenation of technical indicators derived from raw price and volume data. Specifically, we include:

\begin{itemize}
    \item \textbf{Moving Averages:} For a given time series $\{P_t\}$ representing the daily closing price, the $k$-day simple moving average (SMA) at time $t$ is defined as:
    \[
        \text{SMA}^{(k)}_t = \frac{1}{k} \sum_{i=0}^{k-1} P_{t-i}, \quad k \in \{10, 20, 50\}
    \]

    \item \textbf{Momentum Indicators:} Momentum is computed as the price change over a fixed lag $k$, which captures short-term trend dynamics:
    \[
        \text{Momentum}^{(k)}_t = P_t - P_{t-k}, \quad k \in \{5, 10\}
    \]

    \item \textbf{Volatility Estimates:} We estimate historical volatility using the standard deviation of returns over a rolling window:
    \begin{align*}
        \text{Volatility}^{(k)}_t &= \sqrt{ \frac{1}{k-1} \sum_{i=0}^{k-1} \left( r_{t-i} - \bar{r}_t \right)^2 } \\
        r_t &= \log \left( \frac{P_t}{P_{t-1}} \right)
    \end{align*}
    
    where $\bar{r}_t$ is the mean return over the past $k$ days.
\end{itemize}

These engineered features are computed daily and stacked to form input tensors for all models. This unified feature representation ensures that differences in predictive performance are attributed to model expressiveness rather than discrepancies in input data. It also supports generalization across model families, both classical and quantum-enhanced.

\subsection{Training Framework and Reward Design}

Our agent is trained using the Proximal Policy Optimization (PPO) algorithm, adapted to support both classical and quantum-enhanced policy/value networks. The core learning signal is driven by a proxy reward function designed to encourage the prediction of near-future sector leadership. At each time step $t$, the agent selects a portfolio allocation over $S$ sectors. Let $\mathcal{A}_t$ denote the set of sectors chosen by the agent, and $\mathcal{L}_{t+1}^{(N)}$ denote the set of top-$N$ sectors ranked by market capitalization share at time $t+1$.

The reward $r_t$ is assigned based on a simple yet effective logic:
\begin{itemize}
    \item If the chosen sector at time $t$ is among the top-$N$ ($N=10$ in this work) sectors at $t+1$, the agent receives a reward of 1.0.
    \item Otherwise, a small penalty of -0.1 is applied.
\end{itemize}

Formally, the reward is defined as:
\[
    r_t =
    \begin{cases}
        1.0, & \text{if } \mathcal{A}_t \cap \mathcal{L}_{t+1}^{(N)} \neq \emptyset \\
       -0.1, & \text{otherwise}
    \end{cases}
\]

This proxy reward incentivizes the agent to anticipate short-term market dynamics, rewarding correct predictions and discouraging poor ones. Since incorrect predictions accumulate small penalties, the overall episodic reward becomes a floating-point value, reflecting the balance between accurate and inaccurate actions over time.

However, this reward does not directly optimize long-term investment goals such as cumulative return or Sharpe ratio. Consequently, the learned policy may achieve high episodic rewards during training while underperforming in realistic backtests---an alignment gap that we discuss further in Section~\ref{sec:discussion}. Training is performed on historical data from the Taiwan stock market, with a fixed training--backtesting temporal split to simulate out-of-sample generalization. All model variants, including quantum-enhanced versions, are trained under identical conditions using PPO. The PPO objective seeks to maximize a clipped surrogate advantage function:
\[
    \mathcal{L}^{\text{PPO}}(\theta) = \mathbb{E}_t \left[ \min \left( r_t(\theta) {A}_t, \, \text{clip}(\rho_t(\theta), 1 - \epsilon, 1 + \epsilon) {A}_t \right) \right]
\]
where $\rho_t(\theta) = \frac{\pi_\theta(a_t | s_t)}{\pi_{\theta_{\text{old}}}(a_t | s_t)}$ is the probability ratio between the new and old policy, and ${A}_t$ is the generalized advantage estimate (GAE) computed using the value network $V_\phi(s_t)$. In our quantum variants, the policy $\pi_\theta$ and/or value function $V_\phi$ are implemented using variational quantum circuits (VQCs), enabling the exploration of richer policy classes within the same reinforcement learning framework.

We evaluate the trained policies using standard financial metrics, including cumulative return, annualized return, Sharpe ratio, and maximum drawdown.

\section{Experiments and Results}

\begin{table*}[t]
\caption{Final Training Rewards vs. Evaluation Performance Across Models}
\label{tab:final_comparison}
\centering
\begin{tabular}{lcccccc}
\toprule
\textbf{Model} & \textbf{Final Reward} & \textbf{Cumulative Return} & \textbf{Annualized Return} & \textbf{Annualized Volatility} & \textbf{Sharpe Ratio} & \textbf{Max Drawdown} \\
\midrule
LSTM~\cite{hochreiter1997long}          & 2737.00               & 125.33\%                   & 16.90\%                    & 19.25\%                         & 0.88                  & -31.98\%              \\
Transformer~\cite{vaswani2017attention}    & 2795.30               & 124.29\%                   & 16.80\%                    & 19.16\%                         & 0.88                  & -35.31\%              \\
QNN~\cite{mitarai2018quantum}           & 2847.00               & 105.93\%                   & 14.89\%                    & 19.18\%                         & 0.78                  & -32.26\%              \\
QRWKV          & 2653.40               & 103.91\%                   & 14.68\%                    & 19.18\%                         & 0.77                  & -32.75\%              \\
QASA~\cite{chen2025quantum}         & 2883.30               & 93.14\%                    & 13.49\%                    & 19.03\%                         & 0.71                  & -31.73\%              \\
\bottomrule
\end{tabular}
\end{table*}

\subsection{Experimental Setup}

We evaluate our framework on real-world sector-level financial data from the Taiwan stock market, comprising 2,646 listed companies categorized into 47 industry sectors based on financial statement definitions. The dataset spans from April 23, 2007, to June 13, 2025, and includes daily capital share information, stock prices, and engineered features as described in Section III.

For training, we use data from April 23, 2007, to December 31, 2019, over 100 epochs. The testing period covers January 1, 2020, to June 13, 2025. A rolling-window approach is employed to simulate real-time decision-making and ensure temporal consistency. Feature sequences of length 10 are constructed for each industry to feed into the temporal models.

All models are trained using the Proximal Policy Optimization (PPO) algorithm with the following shared hyperparameters: a discount factor $\gamma = 0.99$, PPO clip parameter $\epsilon = 0.2$, entropy coefficient $\beta = 0.01$, batch size of 64, and 10 PPO epochs per training epoch. The actor and critic networks are trained separately with learning rates of $3\times10^{-4}$ and $1\times10^{-3}$, respectively.

We compare five different model backbones for the policy and value networks:

\begin{itemize}
    \item \textbf{Transformer}: 2-layer transformer with 4 attention heads, 128 hidden dimensions, and dropout 0.1.
    \item \textbf{LSTM}: 2-layer LSTM with 128 hidden units and dropout 0.1.
    \item \textbf{QNN}: Quantum neural network with 4 qubits and 2 quantum circuit layers.
    \item \textbf{QRWKV}: Hybrid classical-quantum recurrent model with 4 heads, 4 layers, embedding size 128, and 4-qubit variational quantum layers.
    \item \textbf{QASA}: Quantum-enhanced attention-based model with 2 layers of 4-head self-attention and 2 quantum circuit layers on 4 qubits.
\end{itemize}

The output of each model corresponds to a probability distribution over 48 investment targets (representing 47 sectors plus a dummy class). The agent selects the top-$N$ sectors ($N = 10$) at each time step. The reward function is binary, assigning a reward of 1 if the selected sector appears among the top-10 sectors by capital share in the next time step, and 0 otherwise.

Evaluation metrics include cumulative return (CR), annualized return (AR), annualized volatility (AV), Sharpe ratio (SR), and maximum drawdown (MDD), providing a comprehensive assessment of both raw return and risk-adjusted performance.

\subsection{Performance Comparison}

Table~\ref{tab:final_comparison} summarizes the final training rewards and evaluation performance of all model variants. While the quantum-enhanced models (QNN, QASA, QRWKV) achieved the highest cumulative training rewards—suggesting that they successfully learned to maximize the designed proxy reward—they consistently underperformed the classical models (LSTM, Transformer) in actual investment metrics such as cumulative return, Sharpe ratio, and drawdown.

This table clearly illustrates the core discrepancy between reward maximization and real-world performance, reinforcing our argument that proxy-based reward functions may not accurately reflect investment quality. In particular, QASA achieved the highest final reward (2883.30), yet delivered the lowest cumulative and risk-adjusted return among all models. Conversely, LSTM and Transformer showed more moderate rewards but better alignment with long-term investment outcomes.

\subsection{Reward-Performance Discrepancy}

A key empirical finding in our study is the apparent disconnect between training reward and actual investment performance. Quantum-enhanced models such as QNN and QASA achieve the highest cumulative reward scores during training, indicating successful learning under the defined reward function. However, these models underperform in real-world backtesting metrics. This inconsistency suggests that while quantum models may be highly expressive and capable of fitting the proxy reward signal, they may also be more prone to overfitting and instability. Their learned strategies exploit reward-specific patterns that do not generalize well to real-world portfolio behavior, especially when risk and volatility are considered.

\subsection{Analysis of Quantum Model Behavior}

Among the quantum-enhanced models, QASA exhibited the poorest performance despite achieving a relatively high training reward. This suggests that its quantum attention mechanism may overemphasize spurious correlations in training data or exhibit optimization instability. In contrast, QNN performed slightly better, but still lagged behind classical models.

The similarity in Sharpe ratios and drawdown across models suggests that the primary differentiator is return generation rather than risk control. Classical models appear to have learned more stable allocation strategies that are better aligned with long-term investment performance, despite receiving lower training rewards.

\subsection{Implications}

These results reinforce the importance of aligning reward design with real-world objectives. In financial applications, maximizing a proxy reward does not guarantee practical investment success. Additionally, the results highlight the sensitivity of quantum-enhanced models to reward formulation and data representation. While they offer theoretical advantages, their practical benefits remain contingent on robust encoding, regularization, and reward shaping strategies. We further discuss these limitations and potential improvements in Section~\ref{sec:discussion}.

\section{Discussion}
\label{sec:discussion}

\subsection{Reward-Performance Misalignment}

One of the most significant findings in this study is the observed disconnect between final training rewards and real-world investment performance. Quantum-enhanced models, particularly QNN and QASA, achieved the highest cumulative rewards during training. However, these models underperformed classical architectures in backtested metrics such as cumulative return and Sharpe ratio. This gap arises from a fundamental issue in financial reinforcement learning: the use of \textit{proxy rewards} that do not fully capture the true objectives of investment performance. In our environment, the agent is rewarded based on whether its selected sector appears in the top-$N$ ranked sectors by future capital share. While this proxy is useful for encouraging short-term predictive accuracy, it fails to directly incentivize risk-adjusted returns, low volatility, or drawdown minimization. As a result, quantum models may learn highly specialized strategies that maximize the proxy reward due to their expressive capacity and susceptibility to overfitting but fail to generalize to broader investment objectives. Classical models, on the other hand, appear to implicitly regularize toward smoother, more stable allocation policies, which align better with investor goals.

\subsection{Broader Implications}

The reward-performance gap we observed is not limited to quantum models, it is a known challenge in deep reinforcement learning for finance. However, the fact that quantum models magnify this gap highlights the urgent need to rethink how reward functions and optimization landscapes interact under quantum regimes. Future research should explore whether quantum models can not only match but exceed classical performance in scenarios where the reward signal better reflects long-term investment goals.

\bibliographystyle{ieeetr}
\bibliography{ref}

\appendix
\section{Additional Experiments}
\begin{figure*}
    \centering
    \includegraphics[width=1\linewidth]{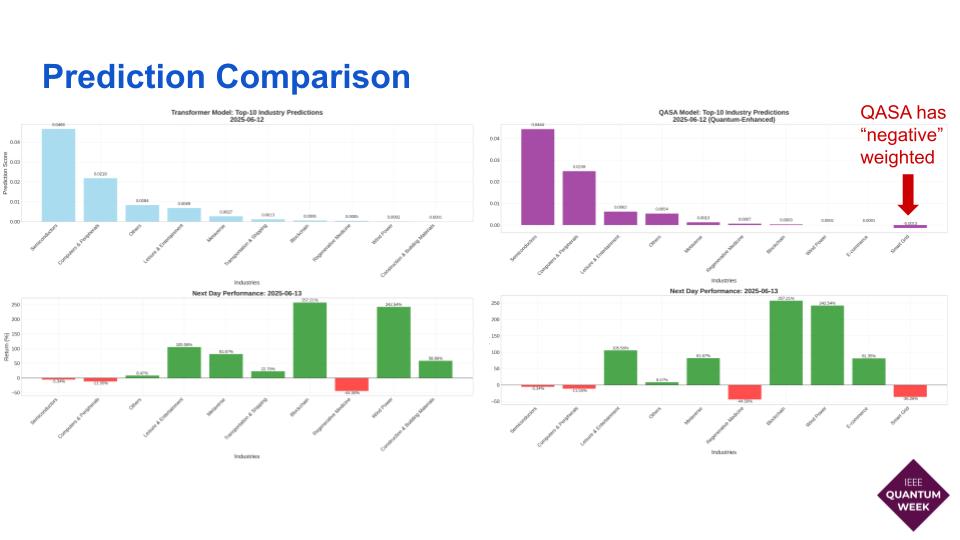}
    \caption{Prediction Comparison}
    \label{fig:pc_qce_ppt}
\end{figure*}

We also compare the prediction result in Fig.~\ref{fig:pc_qce_ppt}.
For institutional investors, QASA’s low drawdown and stable volatility make it more attractive than higher-return but riskier classical baselines.

\end{document}